\documentclass[manuscript]{aastex}

\usepackage[dviout]{graphics/graphicx}
\usepackage{longtable}
\usepackage{lscape}
\usepackage{multirow}
\usepackage{color}
\usepackage{ulem}

\shorttitle{Extinction Laws toward Stellar Sources within a Dusty Circimstellar Medium and Implications for Type Ia Supernovae}
\shortauthors{Nagao et al.}

\begin{document}

\title{Extinction Laws toward Stellar Sources within a Dusty Circimstellar Medium and Implications for Type Ia Supernovae}

\author{Takashi Nagao\altaffilmark{1,4}, Keiichi Maeda\altaffilmark{1,2}, Takaya Nozawa\altaffilmark{3}
}
\altaffiltext{1}{Depertment of astronomy, Kyoto University, Kitashirakawa-Oiwake-cho,
 Sakyo-ku, Kyoto 606-8502, Japan}
\altaffiltext{2}{Kavli Institute for the Physics and Mathematics of the 
Universe (WPI), The University of Tokyo, 
5-1-5 Kashiwanoha, Kashiwa, Chiba 277-8583, Japan}
\altaffiltext{3}{National Astronomical Observatory of Japan, 2-21-1, Osawa, Mitaka, Tokyo 181-8588, Japan}
\altaffiltext{4}{Email: nagao@kusastro.kyoto-u.ac.jp}

\begin{abstract}
  Many astronomical objects are surrounded by dusty environments. In such dusty objects, multiple scattering processes of photons by circumstellar (CS) dust grains can effectively alter extinction properties. In this paper, we systematically investigate effects of multiple scattering on extinction laws for steady-emission sources surrounded by the dusty CS medium, using a radiation transfer simulation based on the Monte Carlo technique. In particular, we focus on whether and how the extinction properties are affected by properties of CS dust grains, adopting various dust grain models. We confirm that behaviors of the (effective) extinction laws are highly dependent on the properties of CS grains. Especially, the total-to-selective extinction ratio $R_{V}$, which characterizes the extinction law, can be either increased or decreased, compared to the case without multiple scattering. We find that the criterion for this behavior is given by a ratio of albedos in the $B$ and $V$ bands. We also find that either small silicate grains or polycyclic aromatic hydrocarbons (PAHs) are necessary for realizing a low value of $R_{V}$ as often measured toward Type Ia supernovae, if the multiple scattering by CS dust is responsible for their non-standard extinction laws. Using the derived relations between the properties of dust grains and the resulting effective extinction laws, we propose that the extinction laws toward dusty objects could be used to constrain the properties of dust grains in CS environments.
\end{abstract} 


\section{Introduction} 
Objects surrounded by the dusty circumstellar (CS) media are often found in the universe, including some classes of supernovae (SNe) (Fox et al. 2011), asymptotic giant branch (AGB) stars, planetary nebulae and protostars. Understanding the properties of the CS dust grains is highly important as they are directly linked to the yet-unresolved host star's evolutionary status. Indeed, different classes of objects are expected to have different components of CS dust grains, depending on the gas composition and chemistry responsible to the creation of those dust grains. We also need to understand a relation between the properties of CS dust grains and the extinction laws they cause, in order to extract intrinsic (dust-free) information on these objects. 

The properties of dust grains are linked to the extinction. It is straightforward to analytically derive the extinction laws caused by interstellar (IS) dust, under the widely adopted assumption that photons which are scattered off by the dust grains do not come back to a line of sight (hereafter a `single scattering' case). This is a reasonable assumption if dust grains are located far away from the source, i.e., IS dust grains. However, in a case of the extinction caused by CS dust grains which are distributed  in the vicinity of the light source, we have to deal with a situation in which photons originally emitted toward a direction out of the line of sight can finally reach an observer after multiple scattering of photons within CS environments (hereafter a `multiple scattering' case). 

Multiple scattering should be in general taken into account in studying objects within dusty CS environments, or inversely in studying the CS environments using the extinction laws. As an example, an increasing attention has been directed to the multiple scattering process within CS environments around Type Ia supernovae (SNe Ia). It has been shown that the extinction laws toward SNe Ia deviate from the standard extinction laws derived for the Milky Way, Large Magellanic Cloud (LMC) or Small Magellanic Cloud (SMC), and show the total-to-selective extinction ratio, $R_{V}$, as small as $\sim 2$ (e.g. Nobili \& Goobar 2008; Folatelli et al. 2010). One possible scenario for explaining this non-standard nature of the extinction law involves the multiple scattering by CS dust grains around SNe Ia (Wang 2005; Goobar 2008). Goobar (2008) showed that the non-standard extinction with a low value of $R_{V}$ can be achieved by the multiple scattering, if CS dust grains have the properties of the MW or LMC dust. He argued that the multiple scattering causes photons with shorter wavelengths to be more efficiently scattered and absorbed than those with longer wavelengths, leading to the steepening of the effective extinction law as compared to the case of the single scattering case. He also pointed out that this scenario requires dust grains with a large albedo, mentioning that the SMC-type dust having a small albedo in the optical to near-IR wavelengths is not suitable for this scenario. However, it was not demonstrated in details how sensitively the effective extinction laws depend on properties of CS dust, and what properties of dust are necessary for realizing unusually low $R_{V}$ through the multiple scattering.

In this paper, we investigate what is a main factor that determines the characteristic properties of the effective extinction laws toward a steady-emission source surrounded by CS dust. In doing this, we adopt various dust models and systematically compute resulting effective extinction laws, by performing Monte Carlo radiation transfer simulations. We especially focus on $R_{V}$, which represents a slope of the extinction curve in the optical regime. In \S 2, we first analytically derive expected relations between an effective extinction law and properties of dust grains under the plane-parallel approximation for the distribution of dust grains. This provides a useful guide for further investing a case of CS dust surrounding the source. In \S 3, we describe our methods for the radiation transfer simulations within the dusty CS environments. In \S 4, we present the results of our simulations. Discussion on relations between dust properties (grain species, size) and values of $R_{V}$ is given in \S 5. In \S 6, we discuss implications for the extinction law toward SNe Ia as an application of the present results, while we emphasize that the general properties we derive in this paper can be applied to any objects surrounded by dusty CS environments. The paper is closed in \S 7 with conclusions.

\section{Extinction through a Plane-Parallel Dusty Cloud}
In this section, we consider an effective extinction law toward a steady light source through a plane-parallel dusty cloud. This subject can be examined in an analytic way. While this situation is not readily applicable to the case of CS dust, the result in this section will provide a useful guide to understand relations between properties of dust grains and a resulting effective extinction law in the case of CS dust. 

\subsection{Values of $R_{V}$ resulting from a semi-infinite plane-parallel dusty cloud}
In this subsection, we derive a value of $R_{V}$ resulting from extinction in a semi-infinite plane-parallel dusty cloud within which dust grains are uniformly distributed. It is assumed that the incident radiation is orthogonal to the plane. Here, two cases are considered. The first is a case in which photons that are scattered off do not come back to a line of sight by another scattering, that is, a case where `multiple scattering' processes are negligible (the `single scattering' case). The second is a case in which this `single scattering' assumption does not hold (the `multiple scattering' case).

In the single scattering case, a value of $R_{V}$ (hereafter $R_{V}^{\mathrm{s}}$) is determined only by the ratio of the mass extinction coefficient of an mixture of dust grains in the $B$ band to the $V$ band ($\kappa_{\mathrm{ext}}(B) / \kappa_{\mathrm{ext}}(V)$)\footnote{The subscript 'ext' is used for values related to 'extinction', i.e., a sum of the contributions from the absorption and scattering.} as follows; 
\begin{equation}
R_{V}^{\mathrm{s}} = \frac{1}{\frac{\kappa_{\mathrm{ext}}(B)}{\kappa_{\mathrm{ext}}(V)}-1} \ .
\end{equation}
Here, $\kappa_{\mathrm{ext}}(B)$ and $\kappa_{\mathrm{ext}}(V)$ include contributions from different dust species, taking into account the size distribution.

In the multiple scattering case, we have to consider photons that are scattered multiple times and finally reach an observer. In this subsection, we assume that a dust grain scatters off photons isotropically, in order to separate effects of anisotropic scattering by dust grains as discussed in \S 4.2. Taking the multiple scattering into account under the plane-parallel approximation, the radiation transfer can be expressed as follows (Chandrasekhar 1960);
\begin{equation}
I_{\lambda}^{1} = I_{\lambda}^{0} \exp [-|k(\lambda)| \tau_{\mathrm{ext}}(\lambda)] \ ,
\end{equation}
where $I_{\lambda}^{0}$ is an incident intensity and $I_{\lambda}^{1}$ is an intensity at the emergence through the plane-parallel cloud. Here, $\tau_{\mathrm{ext}}(\lambda)$ is an optical depth of the cloud along the line of sight at a wavelength $\lambda$. A quantity to characterize the extinction, $k(\lambda)$, is determined by the following equation;
\begin{equation}
\omega(\lambda) = \frac{2k(\lambda)}{\ln\left[\frac{1+k(\lambda)}{1-k(\lambda)}\right]} \ ,
\end{equation}
where $\omega(\lambda)$ is an albedo at a wavelength $\lambda$. We can then obtain $R_{V}$ in the `multiple scattering' case (hereafter $R_{V}^{\mathrm{m}}$) as follows;
\begin{equation}
R_{V}^{\mathrm{m}} = \frac{1}{\frac{k(B) \kappa_{\mathrm{ext}}(B)}{k(V) \kappa_{\mathrm{ext}}(V)}-1} \ .
\end{equation}

If the emission source is embedded in the dusty environment (e.g. CS dust), the effect of the multiple scattering must always be considered. On the other hand, if the scattering clouds are located far away from the source, the importance of the multiple scattering is dependent on a ratio of size of each scattering cloud orthogonal to the line of sight ($L$) to the mean free path in scattering ($r_{0} \sim 1/(\kappa_{\mathrm{scat}}\rho_{0}$), where $\kappa_{\mathrm{scat}}$ is a mass scattering coefficient and $\rho_{0}$ is a typical mass density in the cloud). If $L < r_{0}$ for all the clouds through which the incident radiation propagates, the effects of the multiple scattering should be unimportant. This case applies to extinction by IS dust, and thus $R_{V}^{\mathrm{s}}$ can be used \citep{Spitzer1978}. If $L > r_{0}$ in a specific cloud, the multiple scattering must be taken into account. An exception for the latter is the case where the length scale of scattered echo ($r_{0}$) is spatially resolved, because, in the scale smaller than $r_{0}$, the scattering cloud can be regarded as the single scattering component.

From Equations (3) and (4), it is clear that the value of $R_{V}^{\mathrm{m}}$ depends not only on the ratio of $\kappa_{\mathrm{ext}}(B)$ to $\kappa_{\mathrm{ext}}(V)$, but also on the absolute values of $\omega(B)$ and $\omega(V)$, or equivalently on the ratio of $\omega(B)$ to $\omega(V)$ and the value of  $\omega(B)$. It should be noted that the multiple scattering can be important even when the optical depth along a line of sight is smaller than unity, as long as the optical depth of a direction orthogonal to the line of sight is not negligibly small (i.e., $L > r_{0}$). In plane-parallel dusty clouds, the optical depth of a direction orthogonal to a line of sight is assumed to be infinity. Therefore, $R_{V}$ can be determined from Equation (4), regardless of the optical depth along the line of sight.

Equation (4) and Figure 1 show that $R_{V}^{\mathrm{m}}$ is lower than $R_{V}^{\mathrm{s}}$, when $k(B)/k(V) > 1$, namely when $\omega(B) < \omega(V)$. However, when $\omega(B) > \omega(V)$, $R_{V}^{\mathrm{m}}$ should be higher than $R_{V}^{\mathrm{s}}$. Namely, whether $R_{V}^{\mathrm{m}}$ becomes higher or lower than $R_{V}^{\mathrm{s}}$ depends strongly on the dust properties related to the behavior of albedo as a function of wavelength. As we show below, different dust models predict different ratios of the albedos in the {\it B} and {\it V} bands. 

Figure 2 shows the values of $R_{V}^{\mathrm{m}}$ and $\Delta R_{V}^{1}$ ($= R_{V}^{\mathrm{m}} - R_{V}^{\mathrm{s}}$, difference between $R_{V}^{\mathrm{m}}$ and $R_{V}^{\mathrm{s}}$) for various ratios of the mass extinction coefficient in the $B$ band to that in the $V$ band ($\kappa_{\mathrm{ext}}(B)/\kappa_{\mathrm{ext}}(V)$) and for various ratios of the albedo in the $B$ band to that in the $V$ band ($\omega(B)/\omega(V)$). For the purpose of illustration, two cases for $\omega(B) = 0.6$ and $\omega(B) = 0.7$ are shown in the figure. Figure 2 shows how strongly $R_{V}^{\mathrm{m}}$ depends on parameters of dust models. $R_{V}^{\mathrm{m}}$ is lower when $\kappa_{\mathrm{ext}}(B)/\kappa_{\mathrm{ext}}(V)$ is higher and/or $\omega(B)/\omega(V)$ is lower, if $\omega (B)$ is fixed. On the other hand, given $\kappa_{\mathrm{ext}}(B)/\kappa_{\mathrm{ext}}(V)$ and $\omega(B)/\omega(V)$, $|\Delta R_{V}^{1}|$ is higher for higher $\omega (B)$, which means that as the albedo in the $B$ band becomes higher, the effect of the multiple scattering becomes more significant.

\subsection{Cases of typical dust models}
In this subsection, we apply the general discussion in the former subsection to widely used dust models. For demonstration, we consider four dust models, which enables us to provide a fair comparison to the results by Goobar (2008) and obtain further insight; the MW dust and LMC dust models with a power-law size distribution (hereafter MW1 and LMC1, respectively), and the MW dust and LMC dust models with the size distributions proposed by Weingartner \& Draine (2001; hereafter MW2 and LMC2). MW2 and LMC2 are the dust models used by Goobar (2008). These dust models are constructed as a mixture of astronomical silicate and graphite grains of various sizes (and also PAHs for MW2 and LMC2), assuming that dust grains have a spherical shape. We consider that the power-law size distribution (for MW1 and LMC1) has an index of $-3.5$ and that the maximum and minimum grain radii are $0.25$ micron and $0.005$ micron, respectively (Mathis et al. 1977). The mass ratio between astronomical silicate and graphite is $0.642:0.358$ for the MW1 dust model and $0.843:0.157$ for the LMC1 dust model. These values are determined so that the dust models well reproduce the observed extinction curves in the MW and the LMC.\footnote{The mass ratios between silicate and graphite given by Draine and Lee (1984) and Pei (1992) are $0.64:0.36$ and $0.88:0.12$, respectively, for the MW and LMC dust models. The difference between their values and ours mainly stems from the fact that we adopt the updated extinction curves (Whittet 2003; Gordon et al. 2003) to construct these dust models.} Table 1 summarizes the detailed characteristics of these dust models.

The values of $R_{V}^{\mathrm{s}}$ and $R_{V}^{\mathrm{m}}$ in these models, which are expressed as $R_{V}^{\mathrm{s}}$(pp) and $R_{V}^{\mathrm{m}}$(pp)\footnote{The notation `pp' is from Plane Parallel.} respectively, are shown in Table 1. In the LMC1 and MW1 models, $\omega(B) < \omega(V)$ (see Figure 1) and therefore $R_{V}^{\mathrm{m}}$ is higher than $R_{V}^{\mathrm{s}}$. On the other hand, $R_{V}^{\mathrm{m}}$ is lower than $R_{V}^{\mathrm{s}}$ in LMC2 and MW2 models. It should be emphasized that these models are both extensively used to describe the interstellar extinction within the MW (MW1 and MW2) and within the LMC (LMC1 and LMC2), but when the multiple scattering is important, they show the opposite behavior in $R_{V}^{\mathrm{m}}$. From Table 1, $k(B)/k(V)$ is higher in order of LMC2, MW2, LMC1, MW1 (see also Figure 1).

According to the above discussion, the value of $R_{V}$ is found to highly depend on the properties of dust ($\kappa_{\mathrm{ext}}(B)/\kappa_{\mathrm{ext}}(V)$, $\omega(B)/\omega(V)$ and $\omega (B)$) in the case of extinction through the plane-parallel dusty cloud. In the following sections, we study extinction caused by the CS dust grains around a steady-emission source. In this case, we have to deal with extinction taking into account the multiple scattering process, even if the optical depth is low. Unlike the plane-parallel case, the extinction by the CS dust is not described analytically, thus we calculate the value of $R_{V}^{\mathrm{m}}$ using the Monte Carlo radiation transfer method.

\section{Methods}
We simulate the radiation transfer of photons within the CS dust shell around a steadily emitting source, following the prescriptions used by Goobar (2008). The contribution of re-emitted photons following the absorption of a photon within the CS dust is negligible in the $B$ and $V$ bands, given that the temperature of CS dust is at most $\sim 2000$ K. 

We assume that the CS dust grains are distributed uniformly within a sphere of a radius $R_{\mathrm{CS}}$, where $R_{\mathrm{CS}}$ is measured from the central emitting source. Let $n_{0}(B)$ and $n_{0}(V)$ be the numbers of photons in the $B$-band and $V$-band wavelengths, respectively, which are originally emitted from the central source. The propagation of these photons within the CS environment is calculated by 3D Monte Carlo radiation transfer simulations until they reach $R_{\mathrm{CS}}$ and escape the CS environment or until they are absorbed by the CS dust. For each photon path, an optical depth for each interaction of a photon packet is determined following a probability distribution $\exp(-\tau_{\mathrm{ext}})$. The distance of the flight is then obtained from the equation $\tau_{\mathrm{ext}}(\lambda) = \int \kappa_{\mathrm{ext}}(\lambda) \rho dl$, where $\kappa_{\mathrm{ext}}$ is the mass extinction coefficient of a dust model including both the absorption and scattering, and $\rho$ is the mass density of the CS dust and is taken to be constant. At each interaction, the fate of the photon, absorbed or scattered, is determined following the value of albedo $\omega(\lambda)$, through the Monte Carlo random number generation. If the photon is determined to be absorbed, the calculation of this photon packet is finished. If the scattering takes place, a new direction of the photon is specified following a scattering angle distribution. In \S 4.3, the scattering distribution is assumed to be isotropic and in \S 4.1 and \S 4.2, it is assumed to be given by the following Henyey-Greenstein approximation\citep{Henyey1941};
\begin{equation}
\frac{\mathrm{d}\sigma}{\mathrm{d}[\cos{\theta}]} = \frac{1-g^{2}}{[1+g^{2} -2g\cos{\theta}]^{3/2}} \ ,
\end{equation}
where $g = \langle \cos(\theta) \rangle$ (see Table 1) and $\sigma$ is a scattering cross section toward a scattering angle $\theta$. The scattering is isotropic when $g = 0$, while it is perfect forward scattering when $g = 1$. We denote the numbers of photons in the $B$ and $V$ bands escaping the system without absorption as $n_{1}(B)$ and $n_{1}(V)$, respectively. Finally, $R_{V}^{\mathrm{m}}$ can be calculated as follows;
\begin{equation}
R_{V}^{\mathrm{m}} = \frac{A_V}{E(B-V)} = \frac{\log(n_{1}(V)/n_{0}(V))}{\log(n_{1}(B)/n_{0}(B)) - \log(n_{1}(V)/n_{0}(V))} \ .
\end{equation}
Two million photon packets per band are used in these calculations.

\section{Results}
\subsection{Dependence of $R_{V}^{\mathrm{m}}$ on $E(B-V)$}
The Monte Carlo calculations with various amounts of dust are performed. We adopt four different models for the properties of the CS dust grains; MW1, LMC1, MW2, LMC2 (see Table 1). The scattering angle of a photon adopted in the simulations is computed using the Henyey-Greenstein Approximation.

Figure 3 shows the values of $R_{V}^{\mathrm{s}}$ and $R_{V}^{\mathrm{m}}$ for various values of $E (B-V)$. For $0.2 < E (B-V) < 2.0$, $R_{V}^{\mathrm{m}}$ of each model does not depend strongly on $E (B-V)$, as shown in Figure 3, confirming the result by Goobar (2008). Furthermore, these values of $R_{V}^{\mathrm{m}}$ are different from the values of $R_{V}^{\mathrm{s}}$ of the corresponding models. Figure 3 shows $R_{V}^{\mathrm{m}}$(MW1) $\sim 3.9$, $R_{V}^{\mathrm{m}}$(LMC1) $\sim 3.3$, $R_{V}^{\mathrm{m}}$(MW2) $\sim 2.5$, $R_{V}^{\mathrm{m}}$(LMC2) $\sim 1.6$. For the MW2 and LMC2 dust models that Goobar (2008) used, our results are consistent with their results. We also find that the value of $\Delta R_{V}^{1} = R_{V}^{\mathrm{m}} - R_{V}^{\mathrm{s}}$ depends strongly on dust models. In the following subsections, we discuss how the detailed properties of the dust models affect the value of $R_{V}^{\mathrm{m}}$.

\subsection{Dependence of $R_{V}^{\mathrm{m}}$ on $g= \langle \cos(\theta) \rangle$}
Here we perform the same calculations as in \S 4.1, but by changing the value of $g$ artificially, to investigate how the value of $R_{V}^{\mathrm{m}}$ depends on the scattering angular distribution. Figure 4 shows values of $R_{V}^{\mathrm{m}}$ as a function of $E(B-V)$ calculated using a dust model based on LMC2 but with various values of $g$. To simplify the analysis, we used the same value for $g$ in both the $B$ and $V$ bands, ignoring the wavelength-dependence. $R_{V}^{\mathrm{m}}$ is smallest when the scattering is isotropic ($g=0$) and largest when it is described as a perfect forward scattering ($g=1$).

When $g=1$, the value of $R_{V}^{\mathrm{m}}$ can be evaluated by the formula for the single scattering (Equation (1)), but using $\kappa_{\mathrm{abs}}$ instead of $\kappa_{\mathrm{ext}}$;\footnote{Hereafter, $R_{V}^{\mathrm{m}}(g=1)$ is used for a case of perfect forward scattering. $R_{V}^{\mathrm{m}}$ is used for a case of isotropic scattering unless otherwise mentioned.}
\begin{equation}
R_{V}^{\mathrm{m}}(g=1) = \frac{1}{\frac{\kappa_{\mathrm{abs}}(B)}{\kappa_{\mathrm{abs}}(V)}-1} \ .
\end{equation}
Namely, the scattering does not contribute to the extinction law when $g=1$. In this case, due to the perfect forward scattering, the scattering is considered to have null effect on the attenuation. This is different from a case with $g < 1$, where the multiple scattering brings photons originally emitted toward a direction out of the line of sight to the observer's direction, and thus effectively changes the path lengths depending on an albedo.

Equation (7) can be rewritten in a similar form to equation (4) (that assumes isotropic scattering) as follows;
\begin{equation}
R_{V}^{\mathrm{m}}(g=1) = \frac{1}{\Bigl(\frac{1-\omega(B)}{1-\omega(V)}\Bigr)\frac{\kappa_{\mathrm{ext}}(B)}{\kappa_{\mathrm{ext}}(V)}-1} \ .
\end{equation}
From Equations (1) and (8), it is clear that $R_{V}^{\mathrm{m}}(g=1)$ is lower than $R_{V}^{\mathrm{s}}$ when $\omega(B) < \omega(V)$. We note that the criterion for $R_{V}^{\mathrm{m}} < R_{V}^{\mathrm{s}}$ is indeed the same for the cases with $g=1$ and $g=0$, indicating that this is a general criterion independent from the scattering angular distribution.

Equation (7) provides the following values; $R_{V}^{\mathrm{m}}(g=1;\mathrm{MW1})=4.025$, $R_{V}^{\mathrm{m}}(g=1;\mathrm{LMC1})=3.850$, $R_{V}^{\mathrm{m}}(g=1;\mathrm{MW2})=2.546$ and $R_{V}^{\mathrm{m}}(g=1;\mathrm{LMC2})=1.622$. For the MW2 and LMC2 dust models, these values are close to the values of $R_{V}^{\mathrm{m}}$ with $g$ adopted from the dust models under consideration (i.e., $R_{V}^{\mathrm{m}}$ in the former subsection). From this analysis, we clarify that the result by Goobar (2008) can be understood as a limiting case of the general description of the radiation transfer within the dusty CS environments, which is affected both by the scattering angular distribution and by the ratio of the albedos ($\omega(B)/\omega(V)$). The dust models Goobar (2008) adopted have relatively large values of $g$ (close to the forward scattering), and thus $R_{V}^{\mathrm{m}}$ approaches the above value described mainly by the pure absorptive component (Equation (7)). Their dust models have $\omega(B) < \omega(V)$, leading to $R_{V}^{\mathrm{m}} < R_{V}^{\mathrm{s}}$ (Equation (8)).

\subsection{Dependence of $R_{V}^{\mathrm{m}}$ on $\kappa_{\mathrm{ext}}(B)/\kappa_{\mathrm{ext}}(V)$, $\omega(B)/\omega(V)$, and $\omega(B)$}
In this subsection, we examine dependence of $R_{V}^{\mathrm{m}}$ on $\kappa_{\mathrm{ext}}(B)/\kappa_{\mathrm{ext}}(V)$, $\omega(B)/\omega(V)$, and $\omega(B)$. At the first look, there are four parameters ($\kappa_{\mathrm{scat}}(B)$, $\kappa_{\mathrm{abs}}(B)$, $\kappa_{\mathrm{scat}}(V)$, $\kappa_{\mathrm{abs}}(V)$) which describe the system and determine a value of $R_{V}^{\mathrm{m}}$, except for $g$ discussed in \S4.2. However, once the model is specified by $E(B-V)$, i.e., optical depth, the absolute value of mass absorption coefficient is already absorbed to this parameter. Thus, one can determine the values of $R_{V}^{\mathrm{m}}$ from only three parameters in the dust properties ($\kappa_{\mathrm{ext}}(B)/\kappa_{\mathrm{ext}}(V)$, $\omega(B)/\omega(V)$, and $\omega$(B)), as is the case for the plane-parallel cloud.

To investigate how these three parameters affect $R_{V}^{\mathrm{m}}$, we parameterize $\kappa_{\mathrm{ext}}(B)/\kappa_{\mathrm{ext}}(V)$, $\omega(B)/\omega(V)$, and $\omega(B)$. Once the behaviors are clarified through our analysis, these `input' dust parameters can be compared to any specific physical dust models. The amount of CS dust grains is set so that $E(B-V)$ is between 0.2 and 2.0.

Figure 5 shows the values of $R_{V}^{\mathrm{m}}$ and $\Delta R_{V}^{1}$ as a function of $\kappa_{\mathrm{ext}}(B)/\kappa_{\mathrm{ext}}(V)$ and $\omega(B)/\omega(V)$. We also fix $\omega (B)$ to be $0.6$ or $0.7$, which almost cover the range expected for physical dust models (Table 1). In these calculations, the scattering of photons is assumed  to be isotropic ($g=0$). We find that the value of $R_{V}^{\mathrm{m}}$ strongly depends on the detailed properties ($\kappa_{\mathrm{ext}}(B)/\kappa_{\mathrm{ext}}(V)$, $\omega(B)/\omega(V)$, and $\omega(B)$) of dust grains. We also find that the dependence of $R_{V}^{\mathrm{m}}$ on the detailed properties of dust is qualitatively similar to the case of the plane-parallel dusty cloud. $R_{V}^{\mathrm{m}}$ is lower when $\kappa_{\mathrm{ext}}(B)/\kappa_{\mathrm{ext}}(V)$ is higher and $\omega(B)/\omega(V)$ is lower, if $\omega (B)$ is fixed. $\Delta R_{V}^{1}$ is lower for a higher $\omega(B)$ when $\kappa_{\mathrm{ext}}(B)/\kappa_{\mathrm{ext}}(V)$ and $\omega(B)/\omega(V)$ are fixed.

\section{Dust properties and values of $R_{V}^{\mathrm{m}}$}
In this section, we investigate how the value of $R_{V}^{\mathrm{m}}$ depends on dust properties such as the composition and size, by changing the relative mass ratio of silicate and graphite as well as the power-law index, maximum grain radius, and minimum grain radius in the power-law size distribution (i.e., variants of MW1 and LMC1). The values of $\kappa_{\mathrm{ext}}(B)/\kappa_{\mathrm{ext}}(V)$ and $\omega(B)/\omega(V)$ in the various dust models are computed and are compared with the values of $R_{V}^{\mathrm{m}}$ under the plane-parallel approximation with $\omega(B) = 0.6$ and isotropic scattering. This provides a useful estimate for the following reasons. (1) The general behaviors for the plane-parallel dusty cloud and for the CS dust are similar (\S 2.1 \& \S 4.3). (2) The criterion for $R_{V}^{\mathrm{m}} < R_{V}^{\mathrm{s}}$ or $R_{V}^{\mathrm{m}} > R_{V}^{\mathrm{s}}$ (i.e. a ratio of $\omega(B)/\omega(V)$) is independent from $g$. (3) For the parameter range expected for the typical dust models, the effect of $g$ is not large (Figure 4). We however caution that for detailed analysis, especially for dust models with extreme parameters, one has to take the effects of the dust spatial distribution and the scattering angular distribution into account.
            
Figure 6 presents the dependences of $\kappa_{\mathrm{ext}}(B)/\kappa_{\mathrm{ext}}(V)$ and $\omega(B)/\omega(V)$ on the mass ratio between silicate and graphite, and on the power-law index in the size distribution. Here, the maximum and minimum grain radii are fixed to be $0.25$ and $0.005$ micron. $R_{V}^{\mathrm{s}}$ and $R_{V}^{\mathrm{m}}$ become lower for a steeper size distribution and/or for a larger fraction of silicate grains. For example, when the power-law index is $-3.5$, $R_{V}^{\mathrm{s}} \sim 5$ and $R_{V}^{\mathrm{m}} \sim 4$ for the carbon-rich dust model and $R_{V}^{\mathrm{s}} \sim 2.5$ and $R_{V}^{\mathrm{m}} \sim 2.5$ for the silicate-rich dust model, under the plane-parallel approximation with $\omega(B) = 0.6$ and isotropic scattering.

The cases with various maximum grain radii, while the other parameters (the grain size distribution, the mass ratio between silicate and graphite and the minimum grain radius) are fixed at the values adopted in the MW1 model, are shown in Figure 7. The maximum grain radius was changed from $5.0 \times 10^{-3} \mu$m to $1.0 \times 10^{1} \mu$m. $R_{V}^{\mathrm{s}}$ and $R_{V}^{\mathrm{m}}$ take the minimum values when the maximum grain radius is set to be $\sim7.0 \times 10^{-2} \mu$m. Under the plane-parallel approximation with $\omega(B) = 0.6$ and isotropic scattering, the corresponding minimum values of $R_{V}^{\mathrm{s}}$ and $R_{V}^{\mathrm{m}}$ are $\sim 1.5$ and $\sim 2$, respectively, for a specific range of the maximum grain radius between $5 \times 10^{-2} \mu$m and $8 \times 10^{-2} \mu$m. We also performed the calculations with different minimum grain radii, but we find that $R_{V}^{\mathrm{m}}$ does not depend strongly on the minimum grain radius.

Finally, we consider dust models including PAHs. As demonstration, we convert some fraction of the carbon mass in the dust model into PAHs, while the parameters in terms of size distributions of silicate and graphite are the same as those in the MW1 dust model. The optical properties of PAHs are adopted from Li \& Draine (2001). We adopt 10 \AA\ as a typical size of PAH. Calculated values of $\kappa_{\mathrm{ext}}(B)/\kappa_{\mathrm{ext}}(V)$ and $\omega(B)/\omega(V)$ for various mass ratios of astronomical silicate and graphite (and PAHs) are shown in Figure 8. For a large amount of PAH, $R_{V}^{\mathrm{m}}$ can be much lower than $R_{V}^{\mathrm{s}}$. For example, in the carbon-rich dust model where one third of total mass is in PAHs, $R_{V}^{\mathrm{s}} \sim 3$ and $R_{V}^{\mathrm{m}} \sim 2$ can be realized. In summary, $R_{V}^{\mathrm{m}}$ becomes lower in the dust models that include more silicate grains or smaller grains, and in these cases $R_{V}^{\mathrm{s}}$ becomes lower as well. On the other hand, $R_{V}^{\mathrm{m}}$ can be lower without affecting $R_{V}^{\mathrm{s}}$ ($\sim 3$) substantially in the dust models that include PAHs.

\section{Discussions}
The analysis presented in this paper is generally applicable to a wide range of astronomical objects within dusty environments as long as the dusty region is relatively localized near the source (e.g., CS dust environments). We plan to present several applications in a forthcoming paper (Nagao et al., in prep.), but in this paper we have focused on one specific example -- it is the non-standard extinction law toward SNe Ia. 

SNe Ia show a non-standard extinction law; $R_{V} \lesssim 2$ in contrast to the typical Galactic value of $R_{V} \sim 3$ (e.g. Nobili \& Goobar 2008; Folatelli et al. 2010). As one possible interpretation, it has been claimed that multiple scattering of SN photons by CS dust may lead to the non-standard extinction law. \citet{Goobar2008} presented this argument by calculating the extinction laws toward a steady source within a CS dusty environment, adopting the LMC2 dust model using a Monte Carlo simulation. We reproduce their result that multiple scattering by the LMC2 dust model leads to low $R_{V}^{\mathrm{m}}$. However, we have shown that this result is strongly dependent on the dust properties, and that the multiple scattering can lead to either higher or lower $R_{V}^{\mathrm{m}}$ than the single scattering case. By our analysis, we have found that $R_{V}^{\mathrm{m}}$ is lower for the CS dust with higher $\kappa_{\mathrm{ext}}(B)/\kappa_{\mathrm{ext}}(V)$ and lower $\omega(B)/\omega(V)$.

From our analysis, we emphasize that the multiple scattering model does not necessarily lead to $R_{V}^{\mathrm{m}} < R_{V}^{\mathrm{s}}$. If it is the main process to explain the non-standard value of $R_{V}$ toward SNe Ia, the properties of the CS dust must satisfy particular criteria. Namely, this scenario requires the CS dust having high $\kappa_{\mathrm{ext}}(B)/\kappa_{\mathrm{ext}}(V)$ and low $\omega(B)/\omega(V)$. For example, $\omega(B)/\omega(V) \lesssim 0.97$ to realize a situation that $R_{V}^{\mathrm{s}} \sim 3$ and $R_{V}^{\mathrm{m}} < 2$ when $\omega(B) = 0.7$. The LMC2 and MW2 dust models (Draine 2003, Weingartner \& Draine 2001) adopted by Goobar (2008) have these features, while the LMC1 and MW1 models do not. Inversely, if one has evidence of a sufficient amount of CS dust around SNe Ia from arguments totally independent from the extinction properties (e.g., Maeda et al. 2015), then one could potentially distinguish the dust models, or generally the nature of dust grains that SNe Ia progenitor produced before the explosion. Again, this argument is not restricted to SNe Ia, and applies generally to any sources within dusty CS environments. We have further analyzed the relation between dust properties and resulting $R_{V}^{\mathrm{m}}$. Thereby we have found that small silicate or PAHs within the CS environments is necessary to lead to low $R_{V}^{\mathrm{m}}$. Especially, in the carbon-rich dust model, a large fraction of PAHs, more than a half of graphite mass, are needed to realize a situation that $R_{V}^{\mathrm{s}} \sim 3$ and $R_{V}^{\mathrm{m}} < 2$.

We should note that SNe Ia should be regarded as non-steady sources. In this paper we provide analysis of the CS dust multiple scattering model for SNe Ia using a steady source, under the same assumption as in the previous study by Goobar (2008). In reality, time-dependent effects must be taken into account, including the effect of time evolution of source flux and color (e.g. Amanullah \& Goobar 2011, Krugel 2015 for theoretical studies; Amanullah et al. 2015 for an observational study). In fact, a value of $R_{V}$ is frequently derived around the $B$-band maximum. The echo should have a lower flux and bluer color than the corresponding steady source, because the observed light echo at this moment is caused by the delayed arrival of the radiation that was emitted earlier. Therefore, the effects of the multiple scattering may be smaller than in a steady source. Also, the strength of this effect may be different for different SNe Ia, depending on the early-phase light curve and color evolutions which are more diverse than the maximum phase (Cartier et al. 2011). However, in a case where the scattering is mainly forward scattering (see \S4.2) or the timescale of the echo is small, the effects of the source variability can be ignored and the present results can apply. To clarify how the effects of the multiple scattering contribute to the value of $R_{V}$ toward SNe Ia requires radiation transfer calculations for non-steady sources. We will present such analyses in our forthcoming paper (Nagao et al. in prep), considering the effect of the difference in dust optical properties, taking into account the time evolution.

\section{Conclusions}
Effects of the multiple scattering are important in considering extinction by dust especially when the dust grains are localized around/near the emission source. In a case of extinction by interstellar dust, this effect is generally negligible, except for a case when the size of (spatially-unresolved) scattering bodies, $L$, is larger than the mean free path in scattering, $r_{0}$, in a system dominating the total extinction (e.g. extinction by a very thick molecular cloud). In a case of circumstellar dust, we have to consider such effects even when an optical depth of a scattering cloud is smaller than unity.

In this paper, we have systematically investigated the nature of extinction toward a steadily emitting source within dusty CS environments, by performing Monte Carlo radiation transfer simulations. Especially, we have clarified how $R_{V}$ after the multiple scattering (i.e., as observed) is connected to the optical properties of dust ($\kappa_{\mathrm{ext}}(B)/\kappa_{\mathrm{ext}}(V)$, $\omega(B)/\omega(V)$ and $\omega(B)$) and the physical properties of dust (sizes and species including PAHs). 

Our findings are summarized as follows:
\begin{enumerate}
\item A value of $R_{V}$ after the multiple scattering ($R_{V}^{\mathrm{m}}$) for the extinction by CS dust is almost constant for $0.2<E(\mathrm{B-V})<2.0$, confirming the result by Goobar (2008).
\item $R_{V}^{\mathrm{m}}$ is generally dependent on the scattering angular distribution. The expressions of $R_{V}^{\mathrm{m}}$ in the limiting cases are derived for $g=0$ (Equation (4)) under the plane-parallel approximation, and for $g=1$ (Equations (7) and (8)).
\item A value of $R_{V}^{\mathrm{m}}$ depends strongly on properties of dust. $R_{V}^{\mathrm{m}}$ is lower when $\kappa_{\mathrm{ext}}(B)/\kappa_{\mathrm{ext}}(V)$ is higher and $\omega(B)/\omega(V)$ is lower for given $\omega(B)$.
\item The important optical property of dust as a criterion for $R_{V}^{\mathrm{m}} < R_{V}^{\mathrm{s}}$ (or $R_{V}^{\mathrm{m}} > R_{V}^{\mathrm{s}}$) is the {\it ratio} of the albedos in the $B$ and $V$ bands, not the {\it absolute} values of the albedos. This criterion applies irrespective of the scattering angular distributions.
\item The ratio $\kappa_{\mathrm{ext}}(B)/\kappa_{\mathrm{ext}}(V)$ is higher in the dust models that include more silicate grains or more small-sized grains. On the other hand, $\omega(B)/\omega(V)$ is lower in the dust models that include PAHs.
\item To realize a situation that $R_{V}^{\mathrm{s}} \sim 3$ and $R_{V}^{\mathrm{m}} < 2$ (as an application to the non-standard extinction toward SNe Ia), the CS dust must include PAHs.
\end{enumerate}

\acknowledgments

Data analyses were in part carried out on the PC cluster at Center for Computational Astrophysics, National Astronomical Observatory of Japan. The authors thank Akihiro Suzuki and participants in "Workshop SN MAS 2016" for stimulating discussion. The work has been supported by Japan Society for the Promotion of Science (JSPS) KAKENHI Grant 26800100 (K.M.) and 26400223 (T.Nozawa), and by JSPS Open Partnership Bilateral Joint Research Project between Japan and Chile. The work by K.M.\ is partly supported by World Premier International Research Center Initiative (WPI Initiative), MEXT, Japan.

\clearpage

\begin{figure}
\includegraphics[scale=0.7,angle=-90]{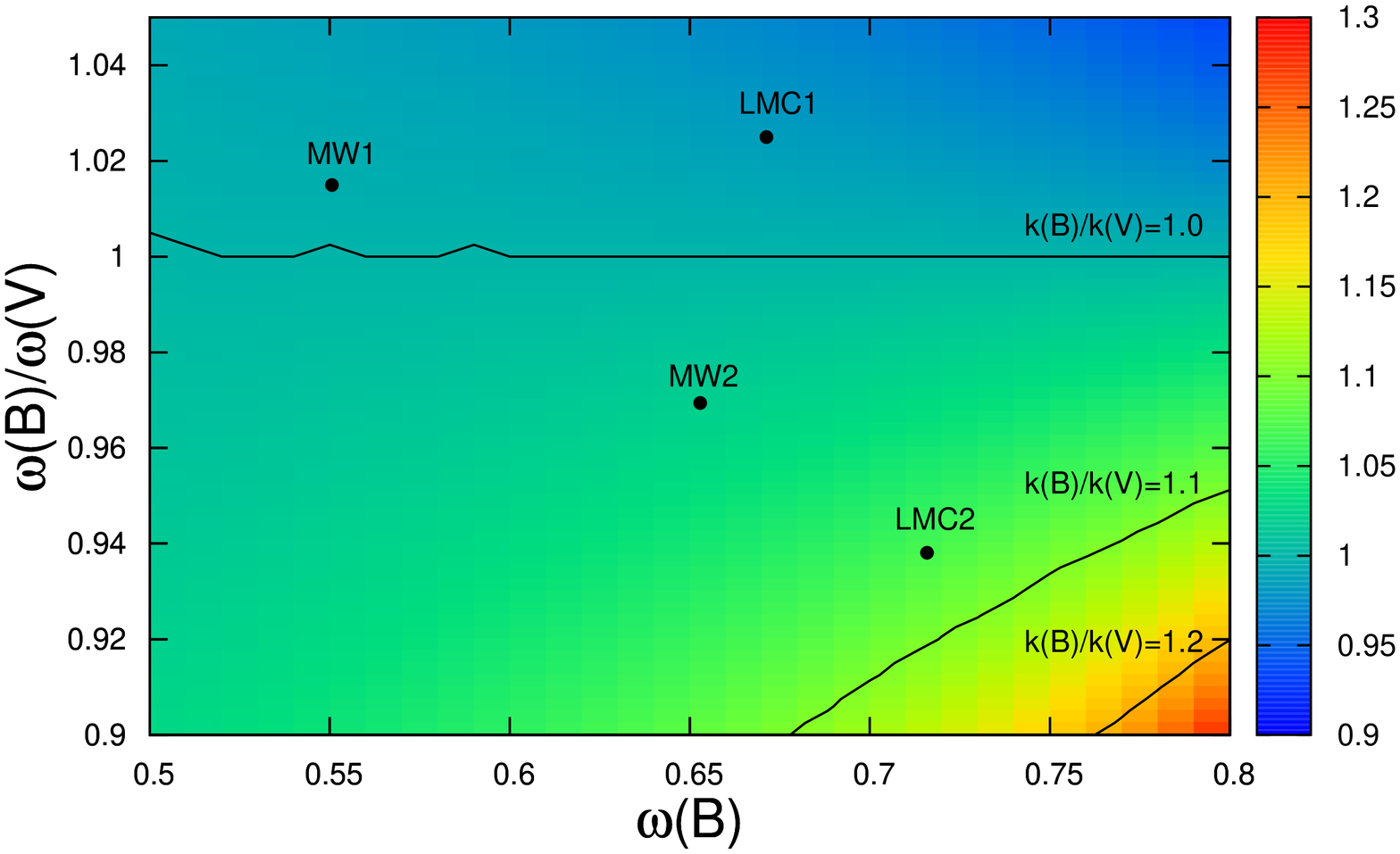}
\caption{Values of $k(B)/k(V)$ (color; see the vertical color bar on the right side for a scale) as a function of $\omega(B)$ and $\omega(B)/ \omega(V)$ (Equation (3)). The properties of the MW1, LMC1, MW2 and LMC2 dust models (see \S2.2) are indicated by black dots. The line of $k(B)/k(V)=1$ provides a criterion for either $R_{V}^{\mathrm{m}} < R_{V}^{\mathrm{s}}$ or $R_{V}^{\mathrm{m}} > R_{V}^{\mathrm{s}}$. The MW1 and LMC1 dust models have the property $k(B)/k(V)>1$ thus $R_{V}^{\mathrm{m}} > R_{V}^{\mathrm{s}}$, while the MW2 and LMC2 dust models have the opposite property.}
\end{figure}

\begin{figure}
\includegraphics[scale=0.5,angle=0]{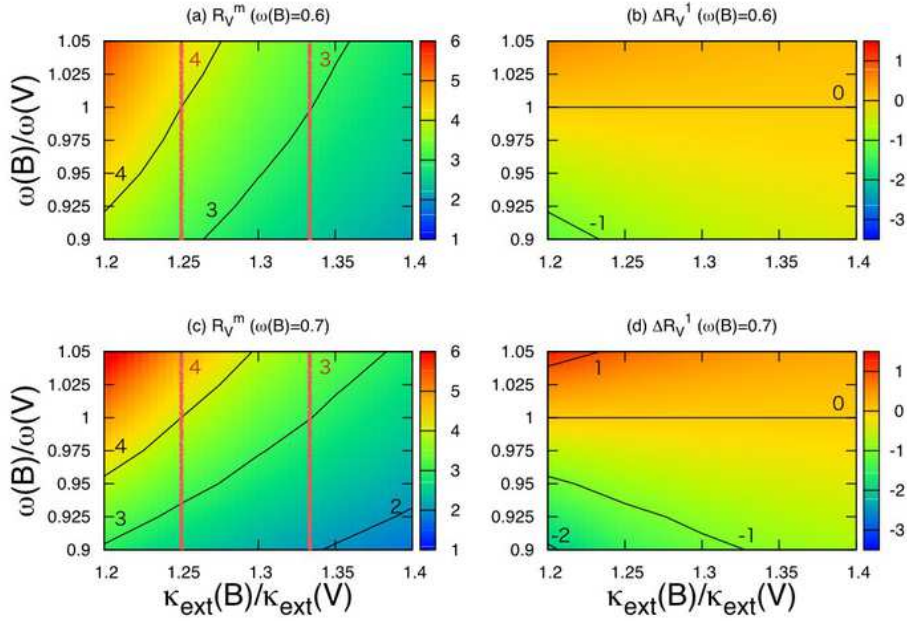}
\vspace{10mm}
\caption{Values of $R_{V}^{\mathrm{m}}$ and $\Delta R_{V}^{1}$($= R_{V}^{\mathrm{m}} - R_{V}^{\mathrm{s}}$) under the plane-parallel approximation with isotropic scattering. The color and the black contour lines show the values ($R_{V}^{\mathrm{m}}$ and $\Delta R_{V}^{1}$) as a function of $\kappa_{\mathrm{ext}}(B)/\kappa_{\mathrm{ext}}(V)$ and $\omega(B)/\omega(V)$ when $\omega(B) = 0.6$ and $0.7$. (a) $R_{V}^{\mathrm{m}}$ for $\omega(B) = 0.6$. (b) $\Delta R_{V}^{1}$ for $\omega(B) = 0.6$. (c) $R_{V}^{\mathrm{m}}$ for $\omega(B) = 0.7$. (d) $\Delta R_{V}^{1}$ for $\omega(B) = 0.7$. The red lines show values of $R_{V}^{\mathrm{s}}$. The criterion for $R_{V}^{\mathrm{m}} < R_{V}^{\mathrm{s}}$ ($R_{V}^{\mathrm{m}} > R_{V}^{\mathrm{s}}$) is given by $\omega(B)/\omega(V)<1$ ($\omega(B)/\omega(V)>1$), and $R_{V}^{\mathrm{m}}$ is smaller for smaller $\omega(B)/\omega(V)$ and/or larger $\kappa_{\mathrm{ext}}(B)/\kappa_{\mathrm{ext}}(V)$.}
\end{figure}

\begin{figure}
\includegraphics[scale=0.7,angle=-90]{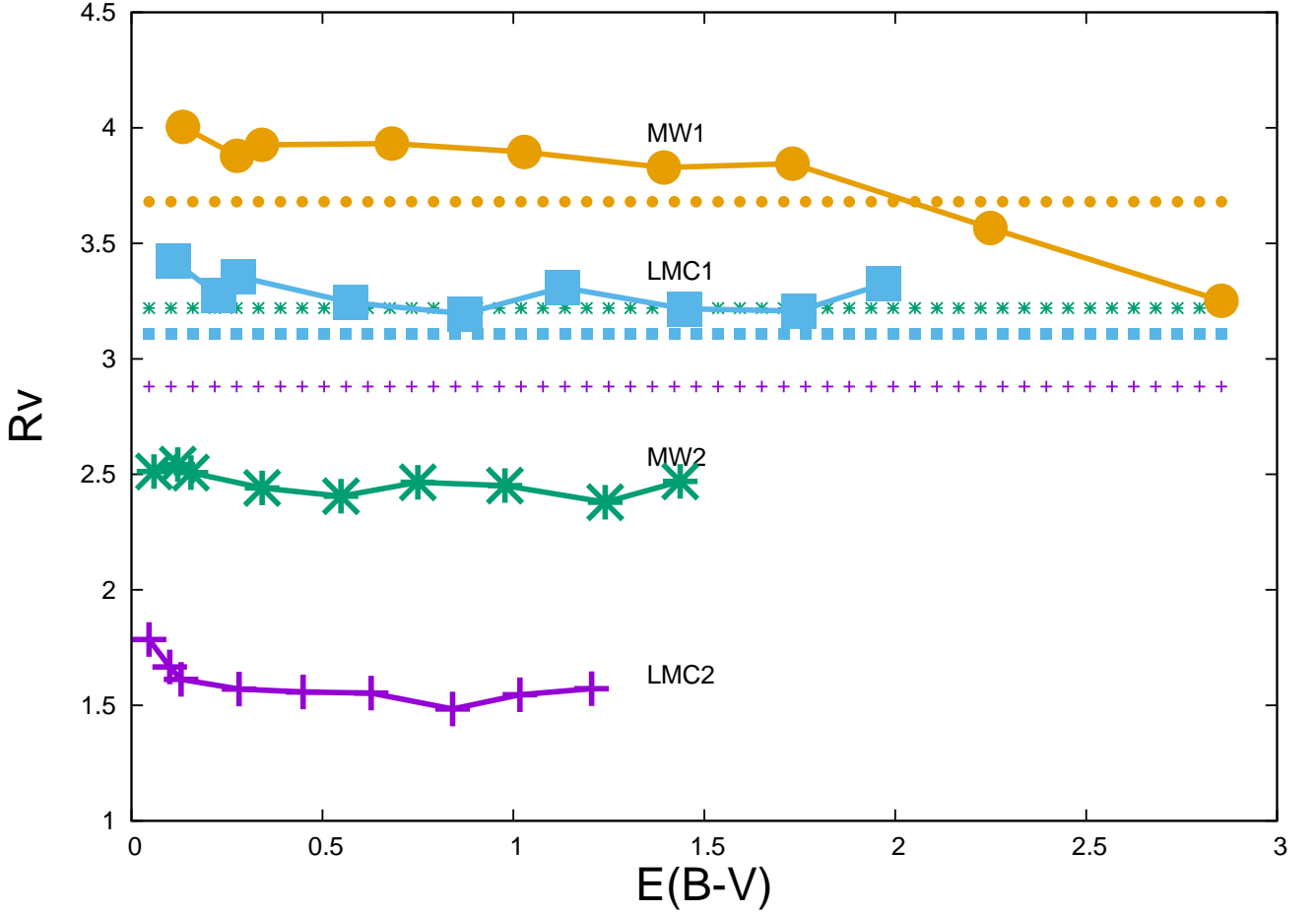}
\caption{$R_{V}^{\mathrm{s}}$ and $R_{V}^{\mathrm{m}}$ as a function of $E(B-V)$ calculated adopting four dust models; MW1 (orange-circles), LMC1 (cyan-squares), MW2 (green-asterisks), and LMC2 (violet-crosses). The values of $R_{V}^{\mathrm{s}}$ for the same dust models are shown by the dashed lines (with the same colors as the corresponding multiple scattering cases). The MW2 and LMC2 dust models result in $R_{V}^{\mathrm{m}} < R_{V}^{\mathrm{s}}$, while it is not the case for the MW1 and LMC1 dust models.}
\end{figure}

\begin{figure}
\includegraphics[scale=0.7,angle=-90]{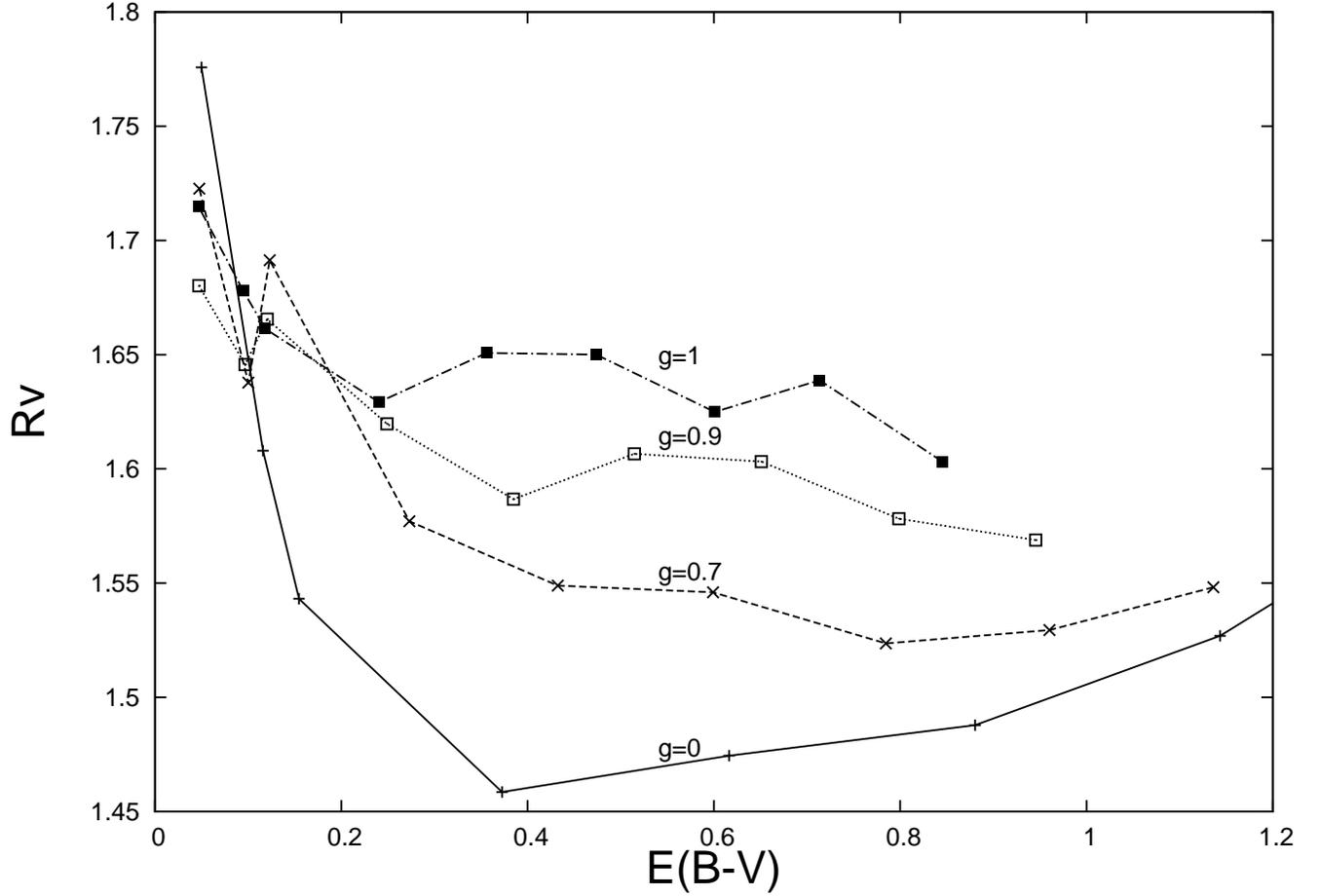}
\caption{$R_{V}^{\mathrm{m}}$ as a function of $E(B-V)$ for various values of $g$ based on the LMC2 model. In these simulations, $g$ is artificially changed from the scattering angular distribution in the original LMC2 dust model. The scattering angular distribution has the effect such that smaller $g$ leads to lower $R_{V}^{\mathrm{m}}$ in this case.}
\end{figure}

\begin{figure}
\includegraphics[scale=0.5,angle=0]{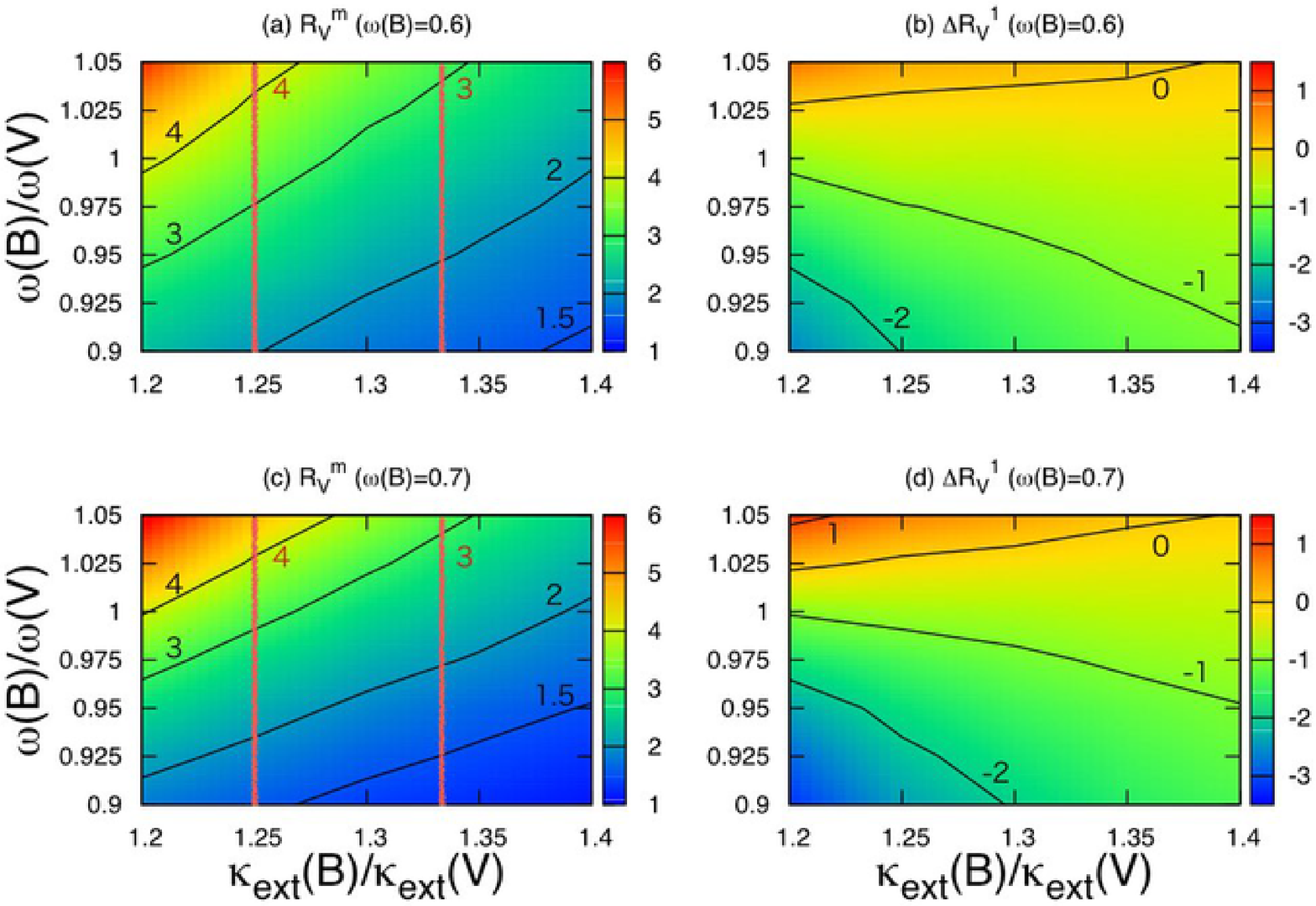}
\vspace{10mm}
\caption{Values of $R_{V}^{\mathrm{m}}$ and $\Delta R_{V}^{1}$($= R_{V}^{\mathrm{m}} - R_{V}^{\mathrm{s}}$) for the case of the CS dust with isotropic scattering. The color and the black contour lines show the values as a function of $\kappa_{\mathrm{ext}}(B)/\kappa_{\mathrm{ext}}(V)$ and $\omega(B)/\omega(V)$ when $\omega(B) = 0.6$ and $0.7$. (a) $R_{V}^{\mathrm{m}}$ for $\omega(B) = 0.6$. (b) $\Delta R_{V}^{1}$ for $\omega(B) = 0.6$. (c) $R_{V}^{\mathrm{m}}$ for $\omega(B) = 0.7$. (d) $\Delta R_{V}^{1}$ for $\omega(B) = 0.7$. The red lines show values of $R_{V}^{\mathrm{s}}$. The general behavior is similar to the case of the plane-parallel dusty cloud (Figure 2).}
\end{figure}

\begin{figure}
\includegraphics[scale=0.5,angle=0]{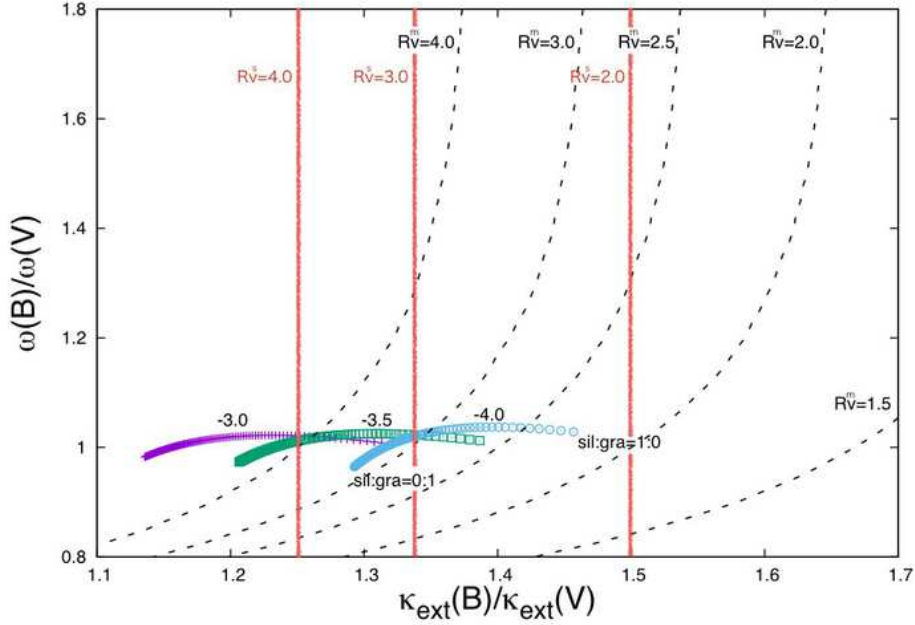}
\caption{The values of $\kappa_{\mathrm{ext}}(B)/\kappa_{\mathrm{ext}}(V)$ and $\omega(B)/\omega(V)$ in the dust models with various mass ratios of astronomical silicate and graphite, and for various size-distributions (assumed to be a power law function in the dust grain radius). Violet-crosses, green-squares, and cyan-circles represent the dust grains where the power-law index in the size distribution is $-3.0$, $-3.5$, and $-4.0$, respectively. For a given power-law index, $\kappa_{\mathrm{ext}}(B)/\kappa_{\mathrm{ext}}(V)$ is larger (to the right) when silicate dust is more abundant. The dashed lines represent contour lines of $R_{V}^{\mathrm{m}}$ in the plane-parallel dusty cloud case (Equation (4)) for $\omega(B) = 0.6$. The red lines show values of $R_{V}^{\mathrm{s}}$. Both $R_{V}^{\mathrm{m}}$ and $R_{V}^{\mathrm{s}}$ become smaller for a dust model with a steeper size distribution and/or a larger abundance of silicate.}
\end{figure}

\begin{figure}
\includegraphics[scale=0.5,angle=0]{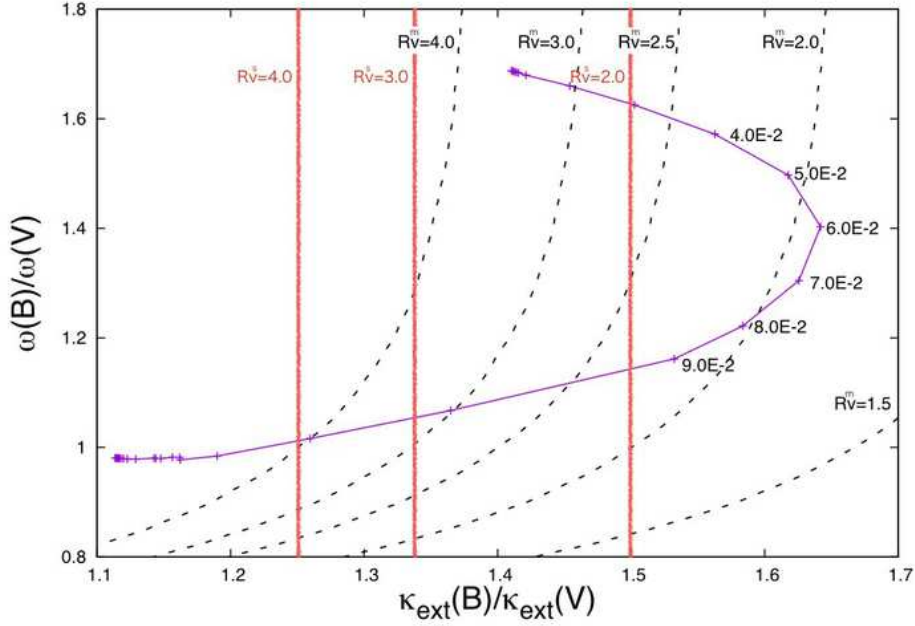}
\caption{The values of $\kappa_{\mathrm{ext}}(B)/\kappa_{\mathrm{ext}}(V)$ and $\omega(B)/\omega(V)$ in the dust models with various maximum grain radii, while the other parameters (the grain size distribution, the mass ratio between silicate and graphite and the minimum grain radius) are fixed at the values adopted in the MW1 model. The maximum grain radius is changed from $5.0 \times 10^{-3} \mu$m to $1.0 \times 10^{1} \mu$m. The dashed lines are the same as in Figure 6. The red lines show values of $R_{V}^{\mathrm{s}}$. The maximum grain radius of (5--8)$\times 10^{-2}$$\mu$m results in $R_{V}^{\mathrm{m}}$ (and $R_{V}^{\mathrm{s}}$) as low as $\sim 2$.}
\end{figure}

\begin{figure}
\includegraphics[scale=0.5,angle=0]{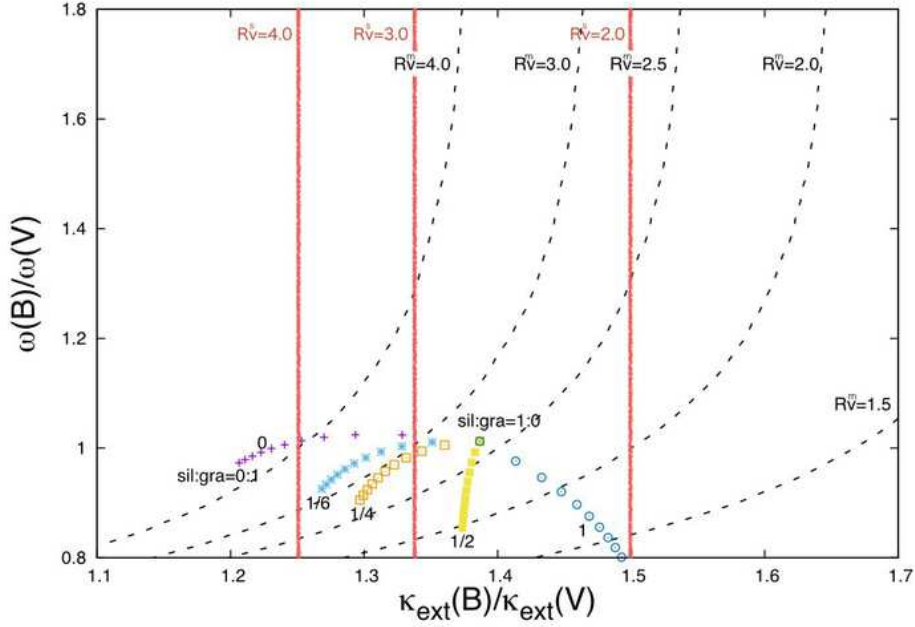}
\caption{The values of $\kappa_{\mathrm{ext}}(B)/\kappa_{\mathrm{ext}}(V)$ and $\omega(B)/\omega(V)$ in the dust models including PAHs, with various mass ratios of astronomical silicate and carbon grains. The mass fraction of PAHs is varied in this figure. Violet-crosses show the case where PAHs are not included. Cyan-asterisk, orange-open-square, yellow-square and blue-circle show the cases where the mass fraction of PAHs is taken as one-sixth, one-fourth, one-second and the same amount, respectively, as compared to the carbon grains. The dashed lines are the same as in Figure 6. The red lines show values of $R_{V}^{\mathrm{s}}$. A dust model including a large amount of PAHs results in $R_{V}^{\mathrm{m}}$ as low as $2$, while $R_{V}^{\mathrm{s}}$ is significantly larger than $R_{V}^{\mathrm{m}}$.}
\end{figure}

\clearpage

\begin{landscape}
\begin{table}
\begin{center}
\caption{Dust parameters}
\begin{tabular}{ccrrrrrrrr}
\tableline
\tableline
Dust model & filter & $\lambda$[$\mu$ m] & $\kappa_{\mathrm{abs}}$[cm$^{2}$/g] & $\kappa_{\mathrm{scat}}$[cm$^{2}$/g] &
 albedo & g & $R_{V}^{\mathrm{s}}$ & $R_{V}^{\mathrm{m}}$(pp) & $\Delta R_{V}^{1}$\\


\tableline
\multirow{2}{*}{MW1}  & B & 0.44 & 1.839E+04 & 2.253E+04 & 0.5506 & 0.4573 & \multirow{2}{*}{3.680} & \multirow{2}{*}{3.759} & \multirow{2}{*}{7.682E-02} \\ \cline{2-7}
                      & V & 0.55 & 1.473E+04 & 1.745E+04 & 0.5423 & 0.4187 & & & \\ \hline
\multirow{2}{*}{LMC1} & B & 0.44 & 9.744E+03 & 1.990E+04 & 0.6713 & 0.5051 & \multirow{2}{*}{3.112} & \multirow{2}{*}{3.320} & \multirow{2}{*}{2.092E-01} \\ \cline{2-7}
                      & V & 0.55 & 7.735E+03 & 1.470E+04 & 0.6552 & 0.4737 & & & \\ \hline
\multirow{2}{*}{MW2}  & B & 0.44 & 1.191E+04 & 2.240E+04 & 0.6529 & 0.5654 & \multirow{2}{*}{3.225} & \multirow{2}{*}{2.973} & \multirow{2}{*}{-2.520E-01}\\ \cline{2-7}
                      & V & 0.55 & 8.551E+03 & 1.764E+04 & 0.6735 & 0.5382 & & & \\ \hline
\multirow{2}{*}{LMC2} & B & 0.44 & 7.542E+03 & 1.900E+04 & 0.7159 & 0.6153 & \multirow{2}{*}{2.876} & \multirow{2}{*}{2.262} & \multirow{2}{*}{-6.139E-01}\\ \cline{2-7}
                      & V & 0.55 & 4.666E+03 & 1.503E+04 & 0.7631 & 0.6059 & & & \\
\tableline
\end{tabular}

\begin{minipage}{.88\hsize}
Notes. The values of MW2 and LMC2 are from Table 1 in \citep{Goobar2008} (The original data is in Weingartner \& Draine (2001) and Draine (2003)).
\end{minipage}
\end{center}
\end{table}
\end{landscape}

\end{document}